\documentclass[preprint,aps,prl,showpacs,superscriptaddress]{revtex4}
\usepackage{amssymb}

\usepackage{graphicx}

\begin{document}
\title{A simple protocol for secure decoy-state quantum key distribution with a loosely controlled source}
\author{Xiang-Bin
Wang\thanks{email: xbwang@mail.tsinghua.edu.cn}\\{Department of
Physics, Tsinghua University, Beijing 100084, China}\\
Cheng-Zhi Peng\thanks{email: pcz@mail.tsinghua.edu.cn}
\\{Department of Physics, Tsinghua University, Beijing 100084,
China}\\{Hefei National Laboratory for Physical Sciences at
Microscale and Department of Modern Physics, University of Science
and Technology of China, Hefei, Anhui 230026, China}\\
Jian-Wei Pan
\thanks{email: jian-wei.pan@physi.uni-heidelberg.de} \\{Department of
Physics, Tsinghua University, Beijing 100084, China} \\{Hefei
National Laboratory for Physical Sciences at Microscale and
Department of Modern Physics, University of Science and Technology
of China, Hefei, Anhui 230026, China}
 \\{Physikalisches Institut, Universit\"at
Heidelberg, Philosophenweg 12, 69120 Heidelberg, Germany}}

\begin{abstract}
The method of decoy-state quantum key distribution (QKD) requests
different intensities of light pulses. Existing theory has assumed
exact control of intensities. Here we propose a simple protocol
which is secure and efficient even there are errors in intensity
control. In our protocol, decoy pulses and signal pulses are
generated from the same father pulses with a two-value attenuation.
Given the upper bound of fluctuation of the father pulses,  our
protocol is secure provided that the two-value attenuation  is done
exactly. We propose to use unbalanced beam-splitters for a stable
attenuation. Given that the intensity error is bounded by $\pm5\%$,
with the same key rate, our method can achieve a secure distance
only 1 km shorter than that of an ideal protocol with exactly
controlled source.
\end{abstract}


\pacs{
03.67.Dd,
42.81.Gs,
03.67.Hk
}
\maketitle


 Recently, some
methods\cite{H03,Wang05,Wang05_2,LMC05,lopra,HQph,scran,kko,zei}
have been proposed for secure quantum key distribution
(QKD)\cite{BB84,GRTZ02,DLH06} with coherent
states\cite{QKD,PNS,PNS1}.  One of these methods is the so called
decoy-state method\cite{H03, Wang05, Wang05_2,LMC05,lopra,HQph}
where Alice randomly changes the intensity of her pulses among a
few values (e.g., 3 values: 0, $\mu$ and $\mu'$) or infinite
values and then she can verify the fraction of single-photon
counts in the raw key. With this information, a secure final key
can be distilled by using the separate theoretical
results\cite{GLLP04}.

So far a number of experiments on decoy-state QKD have been
done\cite{Lo06,peng,ron}, in optical fiber or in free space, in
polarization space or with phase-coding. However, the existing
theory of decoy-state method assumes the exact control of pulse
intensities. A problem met in practice is how to carry out the
decoy-state method efficiently given the inexact control of pulse
intensity. As we have shown\cite{wangnew}, actually, one can
verify the single-photon counts rather efficiently with simple
tomography even though the intensity fluctuations of each light
pulses are large. However, doing it in that way Alice needs
additional operation of tomography.
Here we present a simpler protocol for decoy-state method QKD.

Here we shall consider the effects of inexact control of
$\mu,\mu'$ in a decoy-state protocol requesting 3 intensities,
$\{0,\mu,\mu'\}$. (Actually, all decoy-state protocols with a few
intensities, e.g., 2 or 4 intensities\cite{Wang05_2,lopra} have
the same basic problem of inexact control of $\mu,\mu'$.) For
clarity, lets first consider an {\em ideal protocol} with exact
intensity control, as shown in Fig. 1:a). At any time $t$, if
Alice wants to send a pulse of intensity $\mu$ or $\mu'$, she
first produces a father pulse of intensity $\Omega$. After that
she attenuates the pulse by $A(t)=\mu/\Omega$ or
$A(t)=\mu'/\Omega$ randomly  and a pulse of intensity $\mu$ or
$\mu'$ is produced randomly and sent out to Bob. In this ideal
protocol, Both $\Omega$ and $A(t)$ are controlled exactly.

For practical use, we propose a similar protocol as shown in Fig.
1:b). Alice {\em wants} to produce an intensity $\Omega$ for the
father pulse. She then takes the same random attenuations as that
in the ideal protocol. Here we assume that Alice can control the
attenuation factors of $A(t)$ exactly (either $\mu/\Omega$ or
$\mu'/\Omega$) but she can not control $\Omega$ exactly. ( To
control the instantaneous attenuation $A(t)$ exactly  we can use
unbalanced beam-splitters as we are going to show.) At each time,
she has actually produced intensities of $\{\Omega_t\}$ for the
father pulses. Although we can never control the intensity
exactly, by our currently existing technology, we can definitely
control the intensity in a small range, say, e.g., controlling the
fluctuation within $\pm 5\%$ of $\Omega$. That is to say, Alice
knows the upper-bound of $\{\Omega_t\}$. We denote such an upper
bound value as $\Omega_M$.

Given this upper-bound value $\Omega_M$, the set-up in Fig. 1:b) is
equivalent to a virtual set-up as shown in Fig. 2:a). In the virtual
protocol shown in Fig. 2:a), every time a father pulse of constant
intensity $\Omega_M$ is first produced and then the pulse is
attenuated randomly, with attenuation factors of
$A'(t)=\frac{\Omega_t}{\Omega_M}$. Alice cannot control this
$A'(t)$. After this attenuation, a pulse of intensity $\Omega_t$ is
produced.
 Note that $A'(t)$ is independent of $\mu,\mu'$, since we
can imagine that Alice decides to use $\mu$ or $\mu'$ {\em after}
the attenuation $A'(t)$. Since all attenuators are inside Alice's
Lab., it makes no difference if Alice exchanges the order of
attenuators $A(t)$ and $A'(t)$. The physical meaning of order
exchange is that Alice first decides to use $\mu$ or $\mu'$ and
then arrange the attenuation $A'(t)$ which is independent of
Alice's decision of using $\mu$ or $\mu'$.  This is just the
virtual protocol in Fig. 2:b). In Fig. 2:b), after the pulse
passes $A(t)$ but before passes $A'(t)$, the intensity should be
either $\tilde\mu=\frac{\mu}{\Omega}\Omega_M$ or
$\tilde\mu'=\frac{\mu'}{\Omega}\Omega_M$ {\em exactly}. That is to
say, during the virtual stage between $A(t)$ and $A'(t)$, the
light intensities of each pulses are either exactly $\tilde \mu$
or exactly $\tilde\mu'$. But after a pulse passes through $A'(t)$,
the intensity is changed to inexact values of $\mu_i$ or $\mu_i'$.
For the security proof of the real set-up in Fig. 1:b), we show
the following lemma first. {\bf Lemma:} The set-up in Fig. 2:b) is
unconditionally secure if Alice regards it as a 3-intensity
decoy-state protocol with each light intensities being randomly
chosen from $\{0,\tilde\mu,\tilde\mu'\}$.
\\{\em Proof:} First we suppose Eve controls $A'(t)$.
The dashed square can be regarded as an exact source for a
decoy-state protocol using intensities ${0,\tilde\mu,\tilde\mu'}$.
As it has been known already, decoy-state method with exact
intensity control is secure given whatever channel. Here what Eve
can do is first using $A'(t)$ for attenuation and then do
whatever. This is only a type of specific channel therefore cannot
be used to cheat Alice and Bob. In the set-up of Fig. 2:b),
actually the attenuator $A'(t)$ is not controlled by Eve,
definitely the set-up is secure because Eve cannot attack the
protocol better with her power being reduced. Alternatively, we
imagine that $A'(t)$ is controlled by Alice's friend, Clare. If
Alice choose to disregard Clare's existence then to Alice this is
just a decoy-state protocol with exactly controlled intensities of
$\tilde\mu,\tilde\mu'$.

Given this lemma, we immediately have the {\bf theorem:} {\em The
set-up shown in Fig. 1:b) is unconditionally secure even though
there are intensity fluctuations in values $\mu,\mu'$ if: 1) the
values of $\{\Omega_t\}$ are upper-bounded by $\Omega_M$;  2)
attenuation A(t) is exactly controlled; 3) Alice assumes that she
had used exact intensities of
$\{0,\mu\frac{\Omega_M}{\Omega},\mu'\frac{\Omega_M}{\Omega}\}$ in
calculating the fraction of single-photon counts and distilling
the final key.}

 The proof is simply that the final light pulses produced in Fig. 1:b)  and final light pulses produced
 in Fig. 2:b) are identical. While the scheme in Fig. 2:b) has
 been proven to be secure by our Lemma.

 The question remaining is then how to make stable attenuations
 $A(t)$
 used above. We can realize $A(t)$ by beam-splitters as shown in Fig.(\ref{att}).
 First, we
 attenuate each father pulses ($\Omega_t$) by a fixed attenuation factor of $A_0=
 \frac{\mu+\mu'}{\Omega}$. ($A_0$ can be realized by an unbalanced beam-splitter.)
 Second, after this $A_0$ attenuation,
 we split the beam by a $\mu:\mu'$ beam-splitter and then either the transmitted
 beam $b$ or the reflected beam $b'$ will be blocked and the other one will be
 guided to the optical fiber and sent to Bob.

Suppose the ideal protocol with parameters $\mu_e,\mu_e'$ would
produce a good key rate in a certain experimental condition. Alice
can try to use any intensities around $\mu_e,\mu_e'$ and then
assumes that she had used larger intensities for security.  One
good choice is that Alice tries to produce
$\{0,\mu_e/\lambda,\mu_e'/\lambda\}$ with
$\lambda=\Omega_M/\Omega$ for quantum communication with Bob and
then assumes to have used $\{0,\mu_e,\mu_e'\}$.  For simplicity,
we shall only consider this option hereafter.

The actual efficiency of the protocol can be concluded by a real
experiment. But we can still roughly estimate the efficiency
theoretically on what should be found if we did the experiment.
One can calculate the final key rate\cite{GLLP04,lopra} if he
knows fraction of single-photon counts and the quantum bit error
rate (QBER).

We consider the normal case where there is no Eve. and the channel
transmittance is linear. We shall first compare the key rate of
the following two protocols: our protocol with channel
transmittance $\eta$ (protocol $P(\eta)$) and the ideal protocol
with  channel transmittance $\eta/\lambda$ (protocol
$P_0(\eta/\lambda)$). In both protocols they will assume exact
values $0,\mu_e,\mu_e'$ in the calculation of single-photon counts
and final key distillation. However, in protocol $P(\eta)$, Alice
had done her best to produce $0,\mu_e/\lambda,\mu_e'/\lambda$ for
 quantum communication. In both protocols, Alice and Bob can
find the values of $S_{\mu_e}=1-e^{-\eta\mu_e/\lambda}+d_B,~
S_{\mu_e'}=1-e^{-\eta \mu_e'/\lambda}+d_B$  as the counting rates
(yields) of pulses of intensities (or assumed intensities) of
$\mu_e,\mu_e'$, respectively, where $d_B$ is the dark count rate
of Bob's detector.
    We have the
following joint equations\cite{Wang05} to calculate the
single-photon transmittance for {\em both} protocols:
\begin{eqnarray}\label{fund}
\begin{array}{ll}
e^{-\mu_e}s_0+\mu_e
e^{-\mu_e}s_1+cs_c =S_{\mu_e};\\
e^{-\mu_e'}s_0'+\mu_e' e^{-\mu_e'}s_1'+\left(\frac{
\mu_e'}{\mu_e}\right)^2 e^{\mu_e-\mu_e'}s_c' \le S_{\mu_e'}
\end{array}
\end{eqnarray}
Here $c=1-e^{-\mu_e}-\mu_e e^{-\mu_e}$. Parameters of $s_x$ are
counting rates for states $|x\rangle\langle x|$ from $x$ pulses
($x=0,1$), $s_c$ is the counting rates of state $\rho_c$ (state of
those multi-photon pulses) from $\mu_e$ pulses. Parameters $s_x'$
are counting rates of the same state as defined for $s_x$, but
they are for those states from $\mu_e'$ pulses only. The values of
$s_0,s_0'$ can be deduced from the observed counting rate of those
vacuum pulses. (For the two-intensity protocol\cite{lopra}), one
can assume $s_0=s_0'=0$ for the minimum key rate.) Asymptotically,
$s_x=s_x'$, $s_c=s_c'$. Given a finite number of pulses, $s_x$ and
$s_x'$ can be a bit different but their possible range of
difference can be bounded by classical
statistics\cite{Wang05,Wang05_2,lopra} with exponential certainty
therefore minimum values of $s_1,~s_1'$ can be calculated
numerically. Since $s_1,s_1'$ values for both protocols are
calculated from the same equations with same parameters,  the
value of verified single-photon transmittance ($s_1,s_1'$) of our
protocol $P(\eta)$ is equal to that of the ideal protocol
$P_0(\eta/\lambda)$. This indicates that the two protocols should
have the same fraction of single-photon counts for each intensity.

 We use notation $E,E'$ for the observed error rates of pulses of
 intensity $\mu_e,\mu_e'$, respectively. If they are determined by dark
counts, alignment error and transmission error, these values and
the deduced value of single-photon QBER in our protocol should be
the same with (or a bit less than) those of the ideal protocol
$P_0(\eta/\lambda)$.

 Given all these values requested for the final key distillation  are the same for the
two protocols, we conclude that that the key rate of our protocol
$P(\eta)$ is the same with that of the ideal protocol
$P_0(\eta/\lambda)$. In the case that the intensity fluctuation is
bounded by $5\%$ in our protocol (i.e., $\lambda = 1.05$) and the
light intensity decreases by a half for every 15 kms in both
protocols, the QKD distance of our protocol is only shorter than
that of the ideal protocol by less than 1.06 km if we request the
same key rate for two protocols.  Similar calculation shows that
even the maximum fluctuation is 20\%, the shortened distance is
less than 4 kms. This shows that the secure distance of currently
existing experiments\cite{peng,ron} would keep on exceeding 100
kms if they had used the proposed method here. Since currently
existing experimental results\cite{peng,ron} have not adopted our
protocol (producing different intensities from the same laser
device with attenuation), it should be interesting to redo the
decoy-state QKD experiment using our protocol for a securer
result.

 Acknowledgement. This work was
supported in part by the National Basic Research Program of China
grant No. 2007CB907900 and 2007CB807901,  and China Hi-Tech program
grant No. 2006AA01Z420. We thank Dr Dong Yang, Jun Zhang and
Xian-Min Jin for many helps.

\newpage
\begin{figure}
\centerline{\includegraphics[scale=0.55]{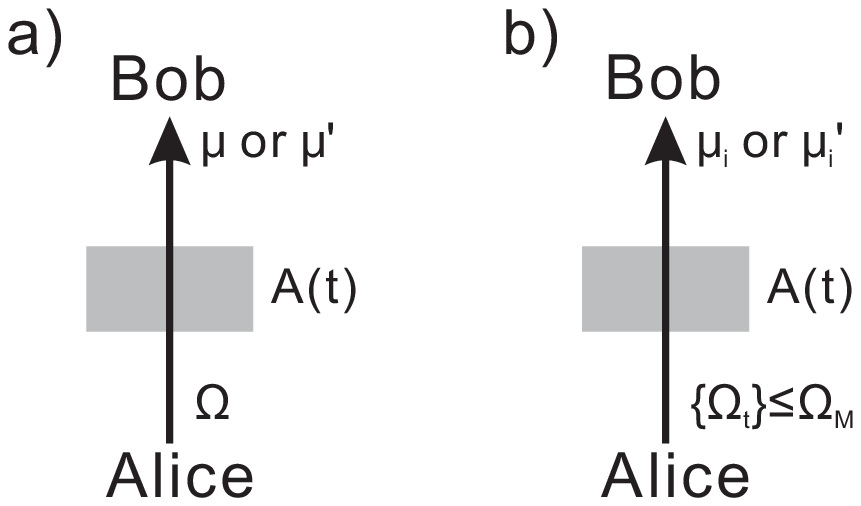}}
\caption{\label{Fig1} a) The ideal protocol that Alice can produce
constant intensity $\Omega$ for the father pulse therefore intensity
$\mu,\mu'$ are controlled exactly. b) The real protocol used in
practice. At each time, Alice {\em wants} to produce intensity
$\Omega$ for the father pulse, however, she actually produces
$\{\Omega_t\}$ at each time $t$. Consequently, the intensities of
output pulses are $\{\mu_i\},\{\mu_i'\}$. We assume that Alice can
control the attenuator $A(t)$ exactly in a real protocol. After a
father pulse is produced, Alice randomly choose the attenuation
factor by $A(t)=\mu/\Omega$ or $A(t)=\mu'/\Omega$. Here the
subscript $t$ is from $1$ to $N+N'$, subscript $i$ for $\{\mu_i\}$
is from $1$ to $N$, subscript $i$ for $\{\mu_i'\}$ is from $1$ to
$N'$.  }
\end{figure}
\begin{figure}
\centerline{\includegraphics[scale=0.65]{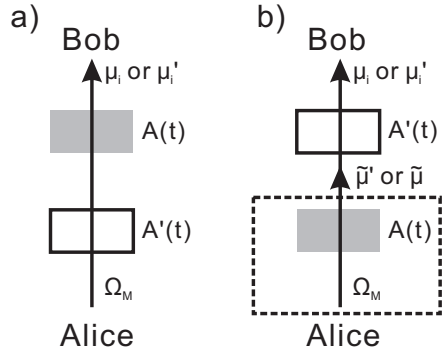}}
\caption{\label{Fig2} Equivalent virtual protocols. Our real
protocol in Fig. 1:b) is equivalent to a virtual protocol as shown
in part a) of this figure. Here Alice first produces a constant
intensity $\Omega_M$ for each father pulses and then attenuates each
of them by attenuator $A'(t)$. After $A'(t)$, the pulse intensity is
$\Omega_t$. It makes no difference to the output light if we
exchange the order of $A(t)$ and $A'(t)$, therefore a) is equivalent
to b). In part b), we can also regard the dashed square as our
source and $A'(t)$ as part of the channel.}
\end{figure}
\begin{figure}
\centerline{\includegraphics[scale=1.0]{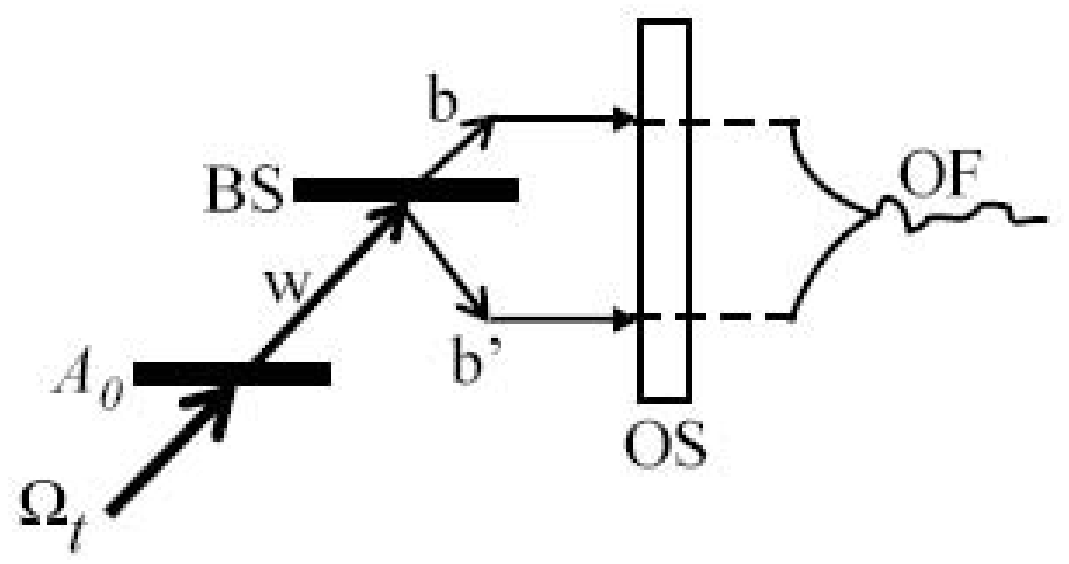}}
\caption{\label{att} Realizing $A(t)$ by unbalanced
beam-splitters. $A_0:$ a constant attenuator with attenuation
factor $\frac{\mu+\mu'}{\Omega}$, this can be realized by an
unbalanced beam-splitter. BS: a $\mu:\mu'$ beam-splitter. OS:
optical switcher. OF: optical fiber.}
\end{figure}

\newpage
\begin{thebibliography}{99}
\bibitem{H03}
W.-Y.~Hwang, Phys. Rev. Lett. {\bf 91}, 057901 (2003).

\bibitem{Wang05}
X.-B.~Wang, Phys. Rev. Lett. {\bf 94}, 230503 (2005).

\bibitem{Wang05_2}
X.-B.~Wang, Phys. Rev. A {\bf 72}, 012322 (2005).

\bibitem{LMC05}
H.-K.~Lo, X.~Ma, and K.~Chen, Phys. Rev. Lett. {\bf 94}, 230504
(2005);
 H.-K. Lo, IEEE ISIT(International Symposium on Information Theory) 2004, p.
137 (IEEE Press. 2004).
\bibitem{lopra}X.~Ma, B. Qi, Y. Zhao, and H.-K. Lo, Phys. Rev. A {\bf 72}, 012326 (2005).
\bibitem{HQph}
J.W.~Harrington {\em et al.}, quant-ph/0503002.
\bibitem{zei} R. Ursin et al, quant-ph/0607182.
\bibitem{scran}
V. Scarani, A. Acin, G. Robordy, N. Gisin, Phys. Rev. Lett. 92,
057901 (2004); C. Branciard, N. Gisin, B. Kraus, V. Scarani, Phys.
Rev. A 72, 032301 (2005).
\bibitem{kko} M. Koashi, Phys. Rev. Lett., 93, 120501(2004); K.
Tamaki, N. L\"ukenhaus, M. Loashi, J. Batuwantudawe,
quant-ph/0608082
\bibitem{BB84}C.H.~Bennett and
G.~Brassard, in {\em Proc.\ of IEEE Int.\ Conf.\ on Computers,
Systems, and Signal Processing (IEEE, New York, 1984)},
pp.~175-179.

\bibitem{GRTZ02}
N.~Gisin, G.~Ribordy, W.~Tittel, and H.~Zbinden,
Rev. Mod. Phys. {\bf 74}, 145 (2002).

\bibitem{DLH06}
M.~Dusek, N.~L\"utkenhaus, M.~Hendrych, "Quantum Cryptography", to
appear in {\em Progress in Optics}, vol. 49, edited by E.~Wolf
(Elsevier, 2006).


\bibitem{QKD}
M.~Bourennane {\em et al.}, F. Gibson, A. Karlsson, A. Hening,
P.Jonsson, T. Tsegaye, D. Ljunggren, and E. Sundberg, Opt. Express
{\bf 4}, 383 (1999); D.~Stucki {\em et al.}, D.~Stucki, N.~Gisin,
O.~Guinnard, G.~Ribordy and H.~Zbinden,  New. J. Physics, {\bf 4},
41, (2002); H.~Kosaka {\em et al.}, Electron. Lett. {\bf 39}, 1199
(2003); C.~Gobby, Z.L.~Yuan, and A.J.~Shields, Appl. Phys. Lett.
{\bf 84}, 3762 (2004); X.-F~Mo {\em et al.}, Opt. Lett. {\bf 30},
2632 (2005); G.Wu, J. Chen, Y. Li, L.-L. Xu and H.-P. Zeng,
quant-ph/0607099.
\bibitem{PNS}
B.~Huttner, N.~Imoto, N.~Gisin, and T.~Mor, Phys. Rev. A {\bf 51},
1863 (1995); H.P.~Yuen, Quantum Semiclassic. Opt. {\bf 8}, 939
(1996)
\bibitem{PNS1}G.~Brassard, N.~L\"utkenhaus, T.~Mor, and
B.C.~Sanders, Phys. Rev. Lett. {\bf 85}, 1330 (2000);
N.~L\"utkenhaus, Phys. Rev. A {\bf 61}, 052304 (2000);
N.~L\"utkenhaus and M.~Jahma, New J. Phys. {\bf 4}, 44 (2002).
\bibitem{GLLP04}
H.~Inamori, N.~L\"utkenhaus, D.~Mayers, quant-ph/0107017;
D.~Gottesman, H.K.~Lo, N.~L\"{u}tkenhaus, and J.~Preskill, Quantum
Inf. Comput. {\bf 4}, 325 (2004).

\bibitem{Lo06}
Y.~Zhao, B. Qi, X. Ma, H.-K. Lo and L. Qian, Phys. Rev. Lett. {\bf
96}, 070502 (2006); Y.~Zhao, B. Qi, X. Ma, H.-K. Lo and L. Qian,
quant-ph/0601168.

\bibitem{peng}  Cheng-Zhi Peng, Jun Zhang, Dong Yang, Wei-Bo Gao, Huai-Xin
Ma, Hao Yin, He-Ping Zeng, Tao Yang, Xiang-Bin Wang and Jian-Wei,
Pan quant-ph/0607129; to appear in Phys. Rev. Lett.

\bibitem{ron} D. Rosenberg, J. W. Harrington, P. R. Rice, P. A. Hiskett, C. G. Peterson,
R. J. Hughes, J. E. Nordholt, A. E. Lita and S. W. Nam,
quant-ph/0607186; to appear in Phys. Rev. Lett.
\bibitem{wangnew} Xiang-bin Wang,
 Phys. Rev. A, 012301(2007), quant-ph/0609081.



\end{thebibliography}
\end{document}